\documentstyle[aps,twocolumn,epsf]{revtex}

\begin{document}
\draft
\title{Vortex lattice in presence of a tunable periodic pinning potential}
\author{W.~V.~Pogosov$^{a}$, A.~L.~Rakhmanov$^{b}$, and V.~V.~Moshchalkov$^{c}$}
\address{$^a$ Moscow Institute of Physics and Technology, 141700\\
Dolgoprudny, Moscow region, Russia\\
$^b$ Institute for Theoretical and Applied Electrodynamics,\\
Russian Academy of Sciences, 127412, Moscow, Russia\\
$^c$ Laboratorium voor Vaste-Stoffysica en Magnetisme, K.U. Leuven,\\
Celestijnenlaan 200 D, B-3001 Leuven, Belgium}
\date{\today}
\maketitle

\begin{abstract}
The vortex patterns stabilized by the square array of artificial pinning
sites with a tunable pinning strength are studied by using a
phenomenological approach in the London limit. The transitions between
pinned and deformed triangular lattices are analyzed as a function of the
amplitude of the vortex-pinning site interaction and the characteristic
length-scale of this interaction. The critical current and different phases
of vortex lattice are studied in presence of external transport current.
\end{abstract}

\pacs{PACS numbers: 74.20.De, 74.60.Ec, 74.25.Ha}

\tighten

\bigskip

\section{Introduction}

Nanoengineered periodic pinning arrays (PPA) have attracted recently a lot
of attention \cite{1,2,3,4,5,6,7,8,9}. By using modern e-beam lithography,
very well defined periodic arrays of submicron antidotes \cite{1,2,4,5,6} or
magnetic dots \cite{7,8,9} can be fabricated in a superconducting film with
a low intrinsic pinning. The latter is needed to have in the film only
artificial periodic pinning centers, which are well controlled. These
laterally nanostructured films demonstrate sharp matching peaks in
magnetization vs. field curves, corresponding to some specific vortex
patterns stabilized by nanoengineered PPA. At temperatures quite close to
the superconducting critical temperature, $T_{c}$, the artificial pinning
sites attract the vortices very strongly and, as a result, the underlying
symmetry of the pinning array dictates the symmetry of stable vortex
patterns. For example, at the matching fields $H=H_{1}$ and $H=H_{1/2}$
(where $H_{1}$ is the first matching field, which generates exactly one flux
quantum per pinning site), square pinning array stabilizes a square vortex
lattice, while in a homogeneous non-patterned reference film a conventional
triangular vortex lattice minimizes the vortex-vortex repulsive interaction.
Stable vortex patterns in presence of the square pinning arrays were
directly visualized with a help of the Lorentz microscopy \cite{6}.

If now the vortex pinning strength in a periodic square array is smoothly
reduced, we can expect that sooner or later the vortex-vortex repulsion will
start to dominate over the pinning forces and the triangular lattice can be
recovered.

In the present paper we have focused on pinning phenomena in presence of a
square array with a tunable pinning strength. We reveal how stable vortex
configurations are evolving as a function of the pinning strength. We have
found the transition lines between pinned and deformed triangular vortex
lattices. At certain matching fields the vortex lattice phase diagram is
more complicated due to the existence of specific periodic phase. We
analyzed the phase diagrams taking into account different types of pinning
potential profile. The critical current in this phase is calculated and the
phase diagrams are obtained.

\section{Phase diagram}

Let us consider the two-dimensional sample with a regular square lattice of
pinning sites with a period $a$. We assume here that the external magnetic
field does not exceed the first matching field $H_{1}$ and is equal to one
of the sub-matching fields. It is well-known that the regular triangular
vortex lattice has the lowest energy in absence of any pinning \cite{10}.
Square lattice of pinning sites can impose its own symmetry on the vortex
lattice. Therefore, vortex lattice can exist in the form of at least two
stable phases - pinned regular phase and triangular phase in two limiting
cases (strong and weak pinning). The symmetry of the pinned lattice is not
necessarily the square one. Vortex patterns corresponding to different
matching fields were listed in Ref. \cite{3} for the limit of strong
pinning. Some of these patterns we also present here on Fig. 1. When the
pinning strength is small it is energetically favorable to form a deformed
triangular lattice. This lattice is nonregular, since triangular vortex
lattice and square pinning site arrays are incommensurate. Transitions
between these phases were first studied in Ref. \cite{11} in the limit of
infinitely small length scale of pinning potential well. Here we will
analyze general case of an arbitrary potential well size.

The pinning energy density can be written phenomenologically as a Fourier
expansion:

\begin{equation}
U(x,y)=\sum_{m,n}A_{mn}\cos \frac{2\pi mx}{a}\cos \frac{2\pi ny}{a}.
\end{equation}
In the pinned regular phase all vortices are located on the pinning sites
and the energy of the system $E_{1}$ (per one vortex) reduces to

\begin{equation}
E_{1}=E_{lat}+U(0,0)=E_{lat}+\sum_{m,n}A_{mn},
\end{equation}
where $U(0,0)$ is the pinning potential in the center of the pinning site
and $E_{lat}$\ is the energy of vortex interaction, which depends on the
symmetry of the flux-line lattice (Fig. 1).

The energy of the deformed triangular lattice $E_{2}$ can be found in the
framework of the elasticity theory, in which a discrete vortex lattice is
replaced by an elastic medium. The accuracy of such an approximation will be
discussed below. The energy $E_{2}$ (per one vortex) is given by

\begin{equation}
E_{2}=E_{tr}+E_{pin}+E_{elast},
\end{equation}
where $E_{tr}$ is the energy of an ideal triangular flux-line lattice, $%
E_{pin}$ is the pinning energy, and $E_{elast}$\ is the energy related to
the lattice distortion. To find $E_{pin}$ and $E_{elast}$ we should
calculate first the deviations of vortices from their positions in
triangular lattice. We introduce two-dimensional vector of deviation of the
elastic medium ${\bf u}(x,y)$. For $E_{pin}$ and $E_{elast}$ we have \cite
{12}:

\begin{equation}
E_{elast}=\frac{S_{0}}{2S}\int dS\left[ \left( c_{11}-c_{66}\right) \left( 
{\bf \nabla u}\right) ^{2}+c_{66}\left( \nabla _{\alpha }u_{\beta }\right)
^{2}\right] ,
\end{equation}
\begin{equation}
E_{pin}=\frac{1}{2S}\int dS\left[ U(x,y)-{\bf u}(x,y){\bf f}(x,y)\right] ,
\end{equation}
where the integration is performed over the area $S$ of the sample, $S_{0}$
is the area of the unit cell of vortex lattice, and ${\bf f}=-dU/d{\bf r}$
is the pinning force acting on a vortex. In Eq. (5) we took into account
only the first term in the energy expansion with respect to ${\bf u}(x,y)$.

The deformation ${\bf u}(x,y)$ can be found by the minimization of the total
energy (3). Using the Fourier representation it is easy to obtain:

\begin{equation}
u_{k\alpha }=-\frac{({\bf kf}_{k})k_{\alpha }}{c_{11}S_{0}{\bf k}^{4}}-\frac{%
{\bf k}^{-2}({\bf kf}_{k})k_{\alpha }-f_{k\alpha }}{c_{11}S_{0}{\bf k}^{2}},
\end{equation}
where

\bigskip ${\bf u}_{k}=\int {\bf u}(r)e^{i{\bf kr}}dS,$ \ \ \ \ ${\bf f}%
_{k}=\int {\bf f}({\bf r})e^{i{\bf kr}}dS.$

Taking into account Eq. (1) one can find the following expressions for ${\bf %
u}(x,y)$: 
\begin{equation}
u_{x}=\frac{a}{2\pi c_{11}S_{0}}\sum_{m,n}A_{mn}\frac{m}{m^{2}+n^{2}}\sin 
\frac{2\pi mx}{a}\cos \frac{2\pi ny}{a},
\end{equation}

\begin{equation}
u_{y}=\frac{a}{2\pi c_{11}S_{0}}\sum_{m,n}A_{mn}\frac{n}{m^{2}+n^{2}}\cos 
\frac{2\pi mx}{a}\sin \frac{2\pi ny}{a}.
\end{equation}
These expressions give the deviations of the vortices from their equilibrium
positions in an ideal lattice. It is should be noted that the term
proportional to $1/c_{66}$ in Eq. (6) disappears in Eqs. (7) and (8). This
term corresponds to twisting of the vortex lattice. Absence of this
contribution in the resulting expressions (7) and (8) implies that {\it %
regular} pinning potential {\it does not twist} the vortex lattice.

Note that within the elasticity theory deviations must be much smaller than
the intervortex distance $d$, which leads to the following constraint:

\begin{equation}
\frac{a}{2\pi c_{11}S_{0}}\sum_{m,n}A_{mn}\frac{n}{m^{2}+n^{2}}\ll d.
\end{equation}

Now we can find $E_{elast}$ and $E_{pin}$ from Eqs. (4) and (5):

\begin{equation}
E_{elast}=\frac{1}{2c_{11}S_{0}}\left( <U^{2}>-<U>^{2}\right) ,
\end{equation}

\begin{equation}
E_{elast}=<U>-\frac{1}{2c_{11}S_{0}}\left( <U^{2}>-<U>^{2}\right) .
\end{equation}
Here we took into account the Parseval's identity

\begin{equation}
\sum_{m,n}A_{mn}^{2}+A_{00}^{2}=2<U^{2}>.
\end{equation}
The elasticity energy defined by Eq. (10) is proportional to the dispersion
of the pinning potential, which is a measure of deviation of function from
its average value and describes its sharpness. The first term in the
right-hand side of Eq. (11) corresponds to the pinning energy in the zero
approximation with respect to ${\bf u}({\bf r})$. In this approximation the
vortex positions are independent of the pinning potential and the elastic
energy equals zero. The last term describes the energy of the correlation
between the vortex positions and the pinning sites. This term is equal
exactly to ${\bf -}E_{elast}$ and it is cancelled in the resulting
expression for the total energy:

\begin{equation}
E_{2}=E_{tr}+<U>.
\end{equation}
Note that Eqs. (11)-(13) are valid for any profile of the pinning potential.
For illustration of our method we show the structure of a deformed
triangular lattice calculated from Eqs. (7) and (8) in Fig. 2 at $H_{1}$, $%
c_{11}S_{0}/U_{00}=0.2$, $\sigma =0.1a$. Open circles denote pinning sites,
black dots - vortices.

Comparing the energy of the pinned lattice $E_{1}$\ with the energy of the
deformed triangular lattice $E_{2}$ we can obtain the criteria for the
transition between them. We choose the Gaussian profile for the
vortex-pinning site interaction:

\begin{equation}
U(x,y)=-U_{0}\exp \left[ -\left( r/\sigma \right) ^{2}\right] ,
\end{equation}
where $U_{0}$\ is amplitude of interaction, $r$ is distance between vortex
and the pinning site, $\sigma $\ and is the potential length scale. The
total pinning potential in the sample is equal to the sum of contributions
from all pinning sites:

\begin{equation}
U(x,y)=-U_{0}\sum_{m,n}\exp \left[ -\frac{1}{\sigma }\left( \left(
am+x\right) ^{2}+\left( an+y\right) ^{2}\right) \right] .
\end{equation}
Taking into account Eq. (15) and the following relation

\mbox{$<$}%
$U(x,y)>=-U_{0}\pi \sigma ^{2}/a^{2}$,

we equate Eqs. (2) and (13) and obtain condition of the transition between
the two vortex phases:

\begin{equation}
U_{0}=\frac{\Delta E}{\pi \frac{\sigma ^{2}}{a^{2}}-\left( \sum_{m=-\infty
}^{\infty }\exp \left( -\frac{a^{2}m^{2}}{\sigma ^{2}}\right) \right) ^{2}},%
\text{ }\Delta E=E_{lat}-E_{tr}.
\end{equation}
This equation defines the boundary between the pinned lattice and the
deformed triangular lattice. The phase diagram is presented in Fig. 3 in the 
$U_{0}-\sigma $ plane. Below the curve 1 the triangular lattice has the
lower energy and upper - the pinned regular phase. The transition between
these phases is discontinuous. It is clearly seen that the pinning
efficiency depends not only on the amplitude of the potential but also on
the size of the potential well. In fact it follows from Eq. (16) that the
pinning efficiency is controlled by parameter

$U_{0}\left[ \pi \frac{\sigma ^{2}}{a^{2}}-\left( \sum_{m=-\infty }^{\infty
}\exp \left( -\frac{a^{2}m^{2}}{\sigma ^{2}}\right) \right) ^{2}\right] $.

The larger is the pinning site well the higher amplitude is required to
stabilize the pinned phase. As we will see below, this phase diagram
corresponds to the vortex lattice at all sub-matching fields not exceeding $%
H_{1}$ ($H_{1/8}$, $H_{1/4}$, $H_{3/8}$, $H_{1/3}$, $H_{2/3}$), except $%
H_{1} $ and $H_{1/2}$, when phase diagrams are different.

At some specific matching fields the structure of vortex lattice can be more
complicated. Below the first matching field, this happens at $H_{1}$ and $%
H_{1/2}$, when pinned lattice has a square symmetry (Fig. 1(d), (e)). In
this case an intermediate phase appears with decreasing pinning strength, in
which vortices in odd rows are depinned and are located in the middle
between pinning sites, see Fig. 4. Hence, only a half of the vortices is
pinned. The existence of this phase was predicted in Ref. \cite{11} at $%
H=H_{1}$ for infinitely small potential well size. The symmetry of such
lattice is close to the triangular one, and the enhancement of the pinning
energy is compensated by the reduction of the vortex interaction energy. We
analyze now the phase diagram at $H=H_{1/2},$ $H_{1}$\ taking into
consideration the existence of this state. Similar to the previous cases the
energy $E_{int}$ of the intermediate phase can be presented as

\begin{equation}
E_{int}=E_{tr}^{\prime }+\frac{1}{2}\left( U(0,0)+U(0,\frac{a}{2})\right) .
\end{equation}
where $E_{tr}^{\prime }$ is the energy of the vortex interaction. Note that $%
E_{tr}^{\prime }$ is close to $E_{tr}$. The boundary between the deformed
triangular lattice and the intermediate state is defined by the following
equation:

\begin{eqnarray}
U_{0} &=&2\left( E_{tr}^{\prime }-E_{tr}\right) \times  \nonumber \\
&&\Biggl[-2\pi \frac{\sigma ^{2}}{a^{2}}+\left( \sum_{m=-\infty }^{\infty
}\exp \left( -\frac{a^{2}m^{2}}{\sigma ^{2}}\right) \right) ^{2}+  \nonumber
\\
&&\sum_{m=-\infty }^{\infty }\exp \left( -\frac{(m+0.5)^{2}}{(\sigma /a)^{2}}%
\right) \times  \nonumber \\
&&\sum_{m=-\infty }^{\infty }\exp \left( -\frac{m^{2}}{(\sigma /a)^{2}}%
\right) \Biggr]^{-1}.
\end{eqnarray}

The boundary between the intermediate phase and the square pinned lattice is
given by:

\begin{eqnarray}
U_{0} &=&2\left( E_{sq}-E_{tr}^{\prime }\right) \times   \nonumber \\
&&\Biggl[\left( \sum_{m=-\infty }^{\infty }\exp \left( -\frac{a^{2}m^{2}}{%
\sigma ^{2}}\right) \right) ^{2}-  \nonumber \\
&&\sum_{m=-\infty }^{\infty }\exp \left( -\frac{a^{2}(m+0.5)^{2}}{\sigma ^{2}%
}\right) \times   \nonumber \\
&&\sum_{m=-\infty }^{\infty }\exp \left( -\frac{a^{2}m^{2}}{\sigma ^{2}}%
\right) \Biggr]^{-1},
\end{eqnarray}

where $E_{sq}$ is the energy of an ideal square vortex lattice ($%
E_{lat}=E_{sq}$).

The vortex phase diagram for $H_{1}$ and $H_{1/2}$ and is presented
schematically in Fig. 5(a). This diagram is more complicated than that shown
in Fig. 3. Below curve 1 the deformed triangular lattice is more favorable
energetically, in the region between curves 1 and 2 - the intermediate
state, and above curve 2 - the pinned square lattice. The transitions
between all these phases are discontinuous. The values of $E_{tr}$, $%
E_{tr}^{\prime }$, and $E_{sq}$ were calculated within the London theory at $%
H=H_{1},$\ $a=\lambda $ for ($\lambda $ is the London penetration depth).
The energy is measured in units of $H_{c}^{2}/8\pi $, where $H_{c}$ is the
thermodynamic critical field.

To analyze how the phase diagram depends on the type of the pinning
potential, we also calculated the diagram for the parabolic pinning well:

\begin{equation}
V(r)=-U_{0}\left[ 1-\left( \frac{r}{\sigma }\right) ^{2}\right] ,\text{ if\ }%
r\leq \sigma \text{; }V(r)=0,\text{ if }r>\sigma ,
\end{equation}
where $r$ is the distance from the center of the well. In this case the
following expressions define the boundaries between the triangular lattice
and intermediate phase, intermediate phase and square lattice, triangular
lattice and square lattice, respectively:

\begin{equation}
U_{0}=\frac{2\left( E_{tr}^{\prime }-E_{tr}\right) }{1-\pi \frac{\sigma ^{2}%
}{a^{2}}},
\end{equation}
\begin{equation}
U_{0}=2\left( E_{sq}-E_{tr}^{\prime }\right) ,
\end{equation}
\begin{equation}
U_{0}=\frac{E_{sq}-E_{tr}^{\prime }}{1-\pi \frac{\sigma ^{2}}{2a^{2}}}.
\end{equation}
The phase diagram for this potential is presented in Fig. 5(b) at $%
a=0.2\lambda $ for $H=H_{1}$ \ There is also the intermediate phase
separating the deformed triangular and the pinned square lattices. The
boundaries are represented by curves 1, 2, and 3. The main difference from
the Gaussian interaction between the vortex and the pinning site is the
existence of the triple point, when the size of the potential well is large.

We can conclude that for any type of the pinning potential the phase diagram
of the vortex lattice at $H_{1}$ and $H_{1/2}$\ in the plane pinning
strength-pinning well size consists of three main regions: the deformed
triangular lattice (most vortices are situated not at the pinning sites),
the ordered intermediate phase (half of vortices are pinned by PPA and half
are interstitials), and the pinned square lattice (all vortices are pinned
by square PPA). However, details of the phase diagram depend on the exact
form of the potential profile. For some types of the pinning potential the
phase diagram has a triple point between these three phases, when the sizes
of the potential wells are large.

\section{Critical current}

In this subsection we study the critical current of the sample in the
intermediate phase. We consider two situations - when the external current
is directed along the $y$- or the $x$-axis on Fig. 4(b). It is obvious that
critical currents are different in these cases in contrast to the pinned
square and the deformed triangular lattices, when they are the same in the $%
x $- and $y$-directions.

In fact there are two systems of the potential wells in the intermediate
phase. The first one is formed by the pinning sites themselves and positions
of these wells are fixed. The second one is formed due to the interaction
with vortices trapped by pinning sites (''caging''). In absence of any
current the second potential wells have minima at the interstitial
positions, see Fig. 4(b). When the transport dc current is applied, the
vortices trapped by pinning sites are displaced, and as a result, the
positions of minima of the potential wells, in which interstitial vortices
are situated, will also change. Therefore, displacements of both
interstitial and pinned vortices depend on each other and on the parameters
of the pinning potential.

Consider first the situation when the external current $j$ flows in the $y$%
-direction and $H=H_{1}$. The Lorentz force acting on vortices is directed
along the $x$-axis and is given by:

\begin{equation}
f_{L}=\frac{2\pi }{\kappa }j.
\end{equation}
Here and below the dimensionless variables are used: the distance $r$, the
magnetic flux density $h$, and the energy density are measured in units of $%
\lambda $, $H_{c}\sqrt{2}$, $H_{c}^{2}/8\pi $, respectively. In this
notation $\Phi _{0}=2\pi /\kappa $, where $\Phi _{0}$ is the magnetic flux
quanta. To calculate the energy of the system we use the usual London
expression for the vortex field \cite{10}:

\begin{equation}
b(r)=\frac{1}{\kappa }K_{0}(r),
\end{equation}
where $K_{0}(r)$ is the modified Bessel function, $r$ is the distance from
the vortex center. The magnetic energy of the vortex is given by \cite{10}:

\begin{equation}
E_{m}=\frac{2\pi }{\kappa }h(0),
\end{equation}
where $h(0)$ is the total magnetic field in the center of vortex. This field
is equal to the sum of fields of each vortex. To describe the pinning
potential we use the Gaussian potential defined by Eq. (15). We denote the
displacements of interstitial and pinned vortices as $x_{1}$ and $x_{2}$.
The energies $F_{1}$ and $F_{2}$ and of these vortices are given by:

\begin{eqnarray}
F_{1} &=&\frac{2\pi }{\kappa ^{2}}\sum_{m,n}K_{0}\left( \sqrt{\left(
a+x_{1}-x_{2}+2am\right) ^{2}+\left( \frac{a}{2}+an\right) ^{2}}\right) 
\nonumber \\
&&-U_{0}\sum_{m,n}\exp \left( -\frac{\left( x_{1}+am\right) ^{2}+\left( 
\frac{a}{2}+an\right) ^{2}}{\sigma ^{2}}\right)  \nonumber \\
&&+f_{L}x_{1},
\end{eqnarray}

\begin{eqnarray}
F_{2} &=&\frac{2\pi }{\kappa ^{2}}\sum_{m,n}K_{0}\left( \sqrt{\left(
a+x_{1}-x_{2}+2am\right) ^{2}+\left( \frac{a}{2}+an\right) ^{2}}\right) 
\nonumber \\
&&-U_{0}\sum_{m,n}\exp \left( -\frac{\left( x_{2}+am\right) ^{2}+\left(
an\right) ^{2}}{\sigma ^{2}}\right)  \nonumber \\
&&+f_{L}x_{2}.
\end{eqnarray}
We omitted all terms independent on $x_{1}$\ and $x_{2}$ in these
expressions. The vortex positions on $x_{1}$\ and $x_{2}$ can be found from
the force balance condition:

\begin{equation}
\frac{\partial F_{1}}{\partial x_{1}}=0,\text{ \ \ \ \ }\frac{\partial F_{2}%
}{\partial x_{2}}=0.
\end{equation}
These equations can be solved numerically. As a result we can find the
critical current, i.e., the maximum value of the current, at which Eqs. (29)
still have solutions. The critical current dependence on the amplitude of
the vortex-pinning site interaction $U_{0}$ is plotted in Fig. 6 at $\sigma
=0.1a,$ $a=\lambda $. Below solid curve the vortex lattice is stable, above
it becomes unstable and vortices start to move.

The situation is more complicated, if the external current flows in the $x$%
-direction, see Fig. 4(b). For this case the system of equations similar to
Eqs. (27)-(29) has been solved. Our results are presented in Fig. 7. We
found that the vortex depinning transition occurs in two steps. First, the
positions of interstitial vortices become unstable at some finite value of
current, and they jump to the neighboring pinning sites in the $y$%
-direction. The dependence of this current on $U_{0}$ is shown in Fig. 7 at $%
\sigma =0.1a,$ \ $a=1$\ (curve 1). The pinned square lattice remains stable
up to the value of critical current (curve 2), which is more than two times
higher than the current leading to the first instability. Thus, the phase
diagram (Fig. 7) consists of three regions: phase with interstitial
vortices, pinned square lattice, and moving vortices. The dependence of the
critical current on $U_{0}$ for this case is also presented in Fig. 6 (dot
curve) for the comparison with the critical current in the $y$-direction
(solid curve). The critical current in the first case is much higher than
that in the second case. This is because for the destruction of the stable
vortex configuration in the first case we need to suppress the potential
wells created by pinning sites. In the second case - only effective
potential wells created by the repulsion with the pinned vortices. Such a
high anisotropy of the critical current is a specific feature of the
intermediate phase, and it can be used for the experimental detection of
this state.

Finally, we note that according to our estimates, based on the results of
Refs. \cite{13,14}, the pinned phases are energetically more favorable in
experiments \cite{2,3,4,5} on the films with antidot arrays, since the
pinning efficiency is very high in such systems. The triangular deformed
lattice and intermediate phase can be studied in the films, where the
pinning strength of the sites can be tuned, for instanse, by using the
periodic arrays of {\it blind} holes.

\section{Conclusions}

We have studied the stable vortex configurations in a regular array of weak
pinning sites at matching fields not exceeding the first matching field $%
H_{1}$. When the pinning efficiency is low enough it is energetically
favorable to have a deformed triangular lattice, when it is sufficiently
high a lattice of pinned vortices is formed. The phase diagrams of the
vortex lattices are calculated in the plane of two coordinates: the
amplitude of vortex-pinning site interaction and the pinning potential well
size. The vortex displacements in the deformed triangular lattice with
respect to the positions in an ideal one are found within the elasticity
theory. When the external field is equal to the first matching field or is
twice lower, the existence of the intermediate regular phase is also
energetically favorable, in which half of the vortices is pinned by periodic
pinning array and the vortex lattice is close to the triangular one. We
analyzed the phase diagrams for this case taking into account different
types of the pinning potential profile. We found that for some types of the
potential the triple point can exist on the phase diagram, where deformed
triangular lattice, pinned square lattice, and intermediate phase coexist.
The critical current in the intermediate phase is calculated. This current
turns out to be highly anisotropic. When current flows parallel to the
pinned vortices rows the depinning occurs when the Lorentz force suppresses
the effective potential wells, in which caged interstitial vortices are
trapped. When current flows perpendicular to the pinned vortices rows, first
these effective ''caging'' potential wells are destroyed, and interstitial
vortices jump to the neighboring vacant pinning sites. The phase diagrams
for the stable vortex lattice in the plane of applied current-pinning
potential amplitude are also obtained.

\section{Acknowledgements}

W. V. P. and A. L. R. acknowledge support by RFBR, Grants Nos. 00-02-18032
and 02-02-06560, by INTAS, Grants Nos. 01-2282 and YSF 01/2-58, and by the
Russian State Program ''Fundamental Problems in Condensed Matter Physics''.
The work in Leuven is supported by the IUAP, GOA, FWO-V, and ESF VORTEX
programs.

\onecolumn

\section*{FIGURE CAPTIONS}

FIG. 1. Vortex patterns corresponding to different matching fields: (a) - $%
1/8H_{1}$ , (b) - $1/4H_{1}$, (c) - $1/3H_{1}$, (d) - $1/2H_{1}$, (e) - $%
H_{1}$. Open circles show positions of the pinning sites, black dots -
vortices.

FIG. 2. The structure of the triangular vortex lattice, deformed by the
square pinning array of Gaussian potential wells, defined by Eq. (20), at $%
H=H_{1}$, $c_{11}S_{0}/U_{0}=0.2$, $\sigma =0.1a$.

FIG. 3. Phase diagram in the plane of the vortex-pinning site interaction
amplitude and the pinning potential length-scale at sub-matching fields not
exceeding $H_{1}$, except and ($H_{1/8}$, $H_{1/4}$, $H_{3/8}$, $H_{1/3}$, $%
H_{2/3}$). Below curve 1 the deformed triangular lattice has the lowest
energy, above -pinned phase.

FIG. 4. The structure of the intermediate phase at $H_{1/2}$ (a) and $H_{1}$%
\ (b). Black dots and open circles denote the positions of the vortices and
the pinning sites. Dashed lines show the symmetry of the vortex lattice.

FIG. 5. Phase diagram in the plane of vortex-pinning site interaction
amplitude $U_{0}$ and the pinning potential length-scale $\sigma $ at $%
H=H_{1}$. Fig. 5(a) corresponds to the Gaussian potential of the
vortex-pinning site interaction, (b) - to the parabolic one. The similar
phase diagrams are also valid for $H=H_{1/2}$.

FIG. 6. Phase diagram of the intermediate phase in the plane: applied
current - amplitude of the vortex-pinning site interaction at $H=H_{1}$, $%
a=1 $, $\sigma =0.1a$. The current flows along the $y$-direction in Fig.
4(b). Solid curve (critical current) shows the boundary between the stable
lattice with the interstitial vortices and the moving vortices. Also shown
for the comparison (dot curve) is the critical current in the $x$-directions
in Fig. 4(b).

FIG. 7. Phase diagram of the intermediate phase in the plane: applied
current - amplitude of the vortex-pinning site interaction $H=H_{1}$, $a=1$, 
$\sigma =0.1a$. The current flows along the $x$-direction. Curve 1
corresponds to the transition from the lattice with interstitial vortices to
the pinned square lattice, curve 2 - to the depinning transition.

\newpage 
\begin{figure}[tbp]
\epsfxsize= .85\hsize
\centerline{ \epsffile{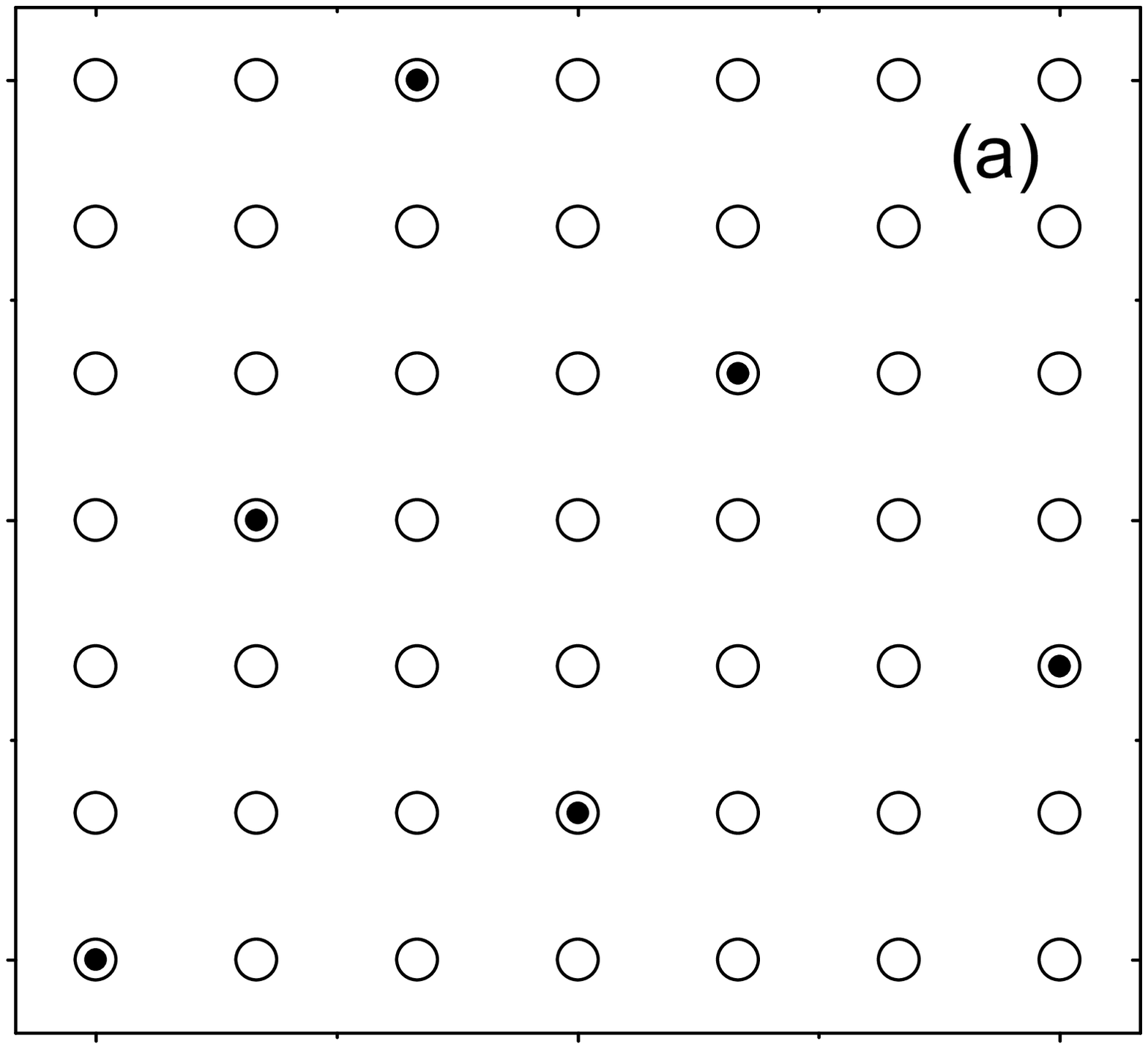}}
\text{Fig. 1(a)}
\epsfxsize= .85\hsize
\centerline{ \epsffile{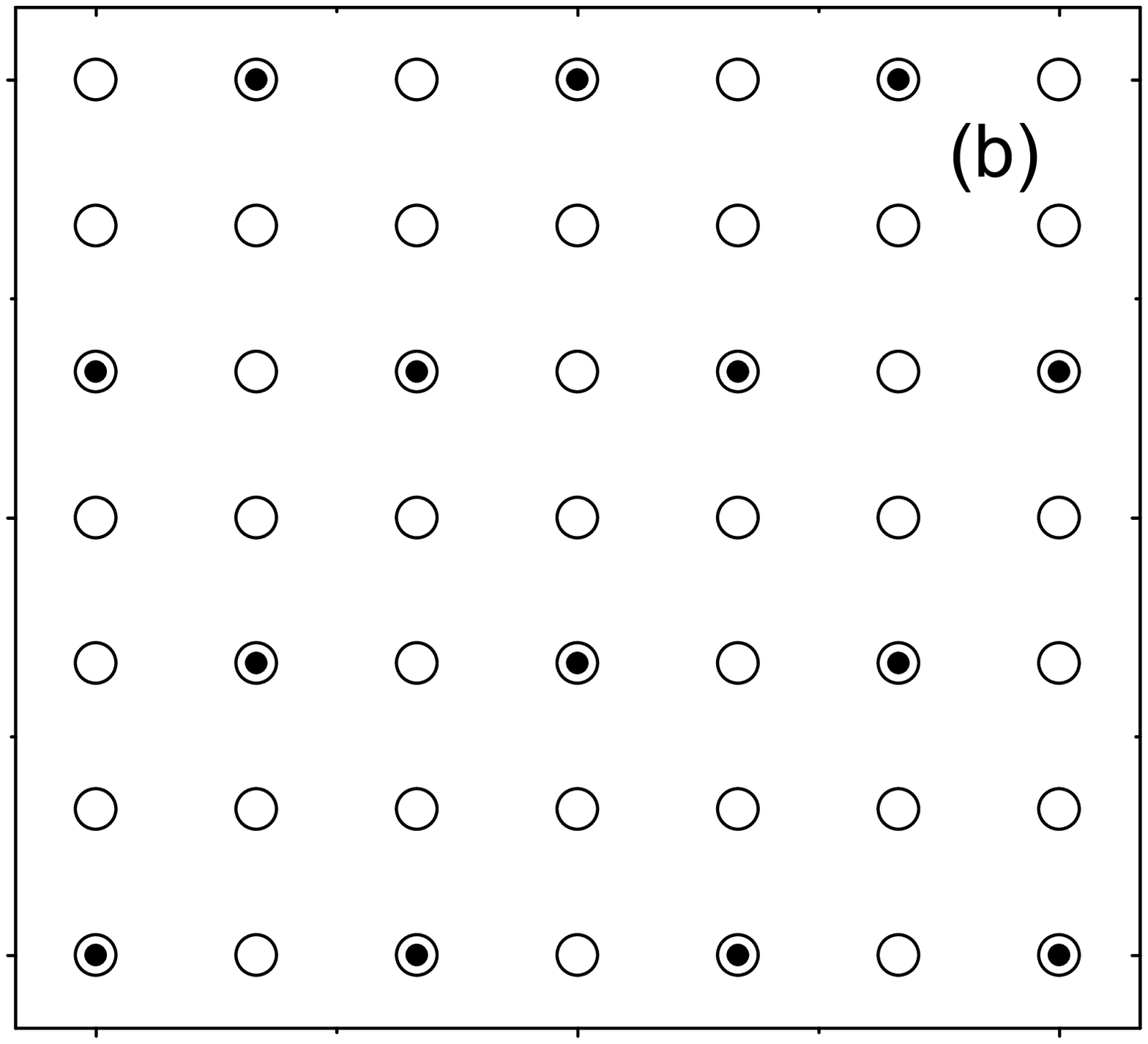}}
\text{Fig. 1(b)}
\newpage
\epsfxsize= .85\hsize
\centerline{ \epsffile{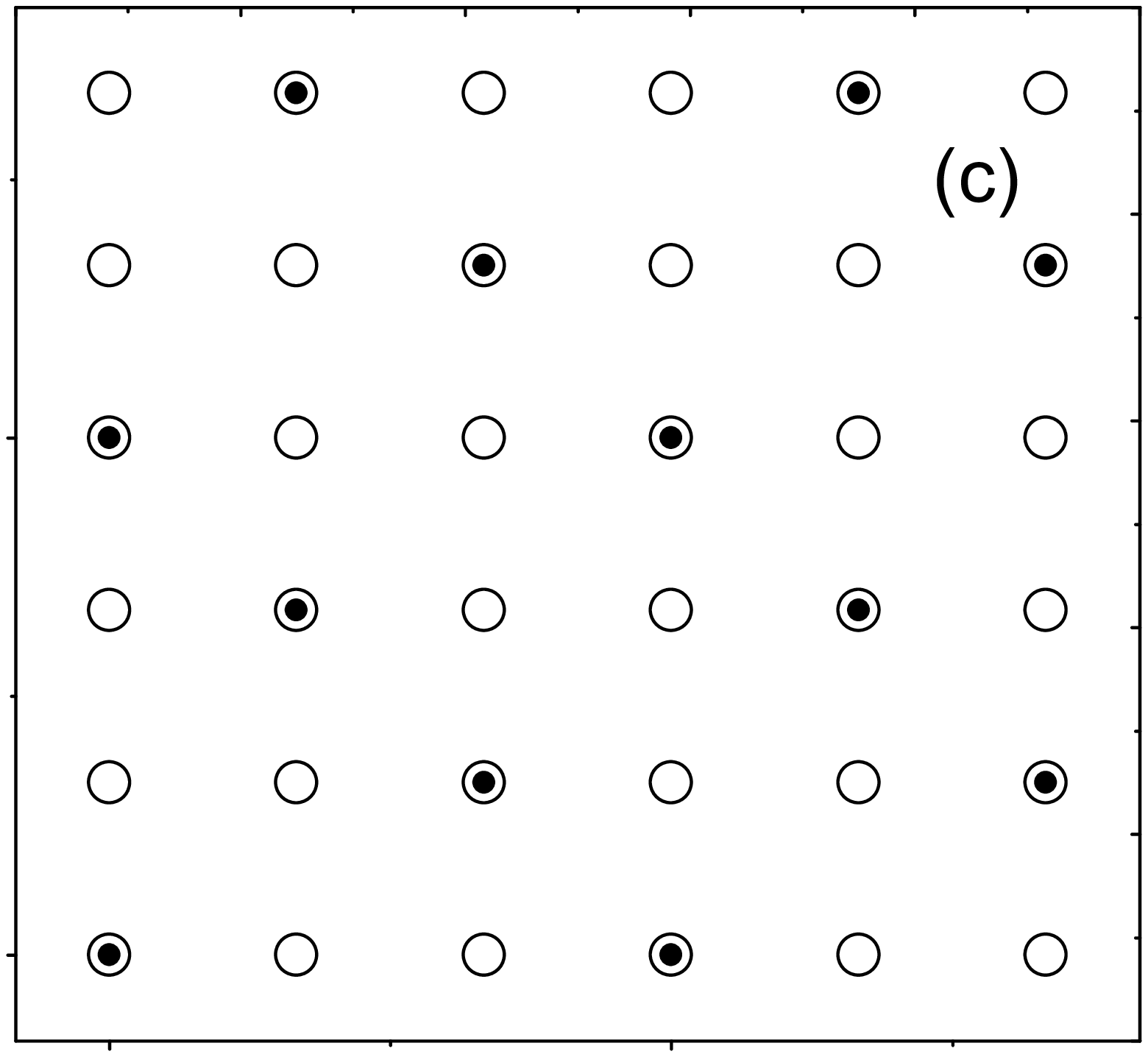}}
\text{Fig. 1(c)}
\epsfxsize= .85\hsize
\centerline{ \epsffile{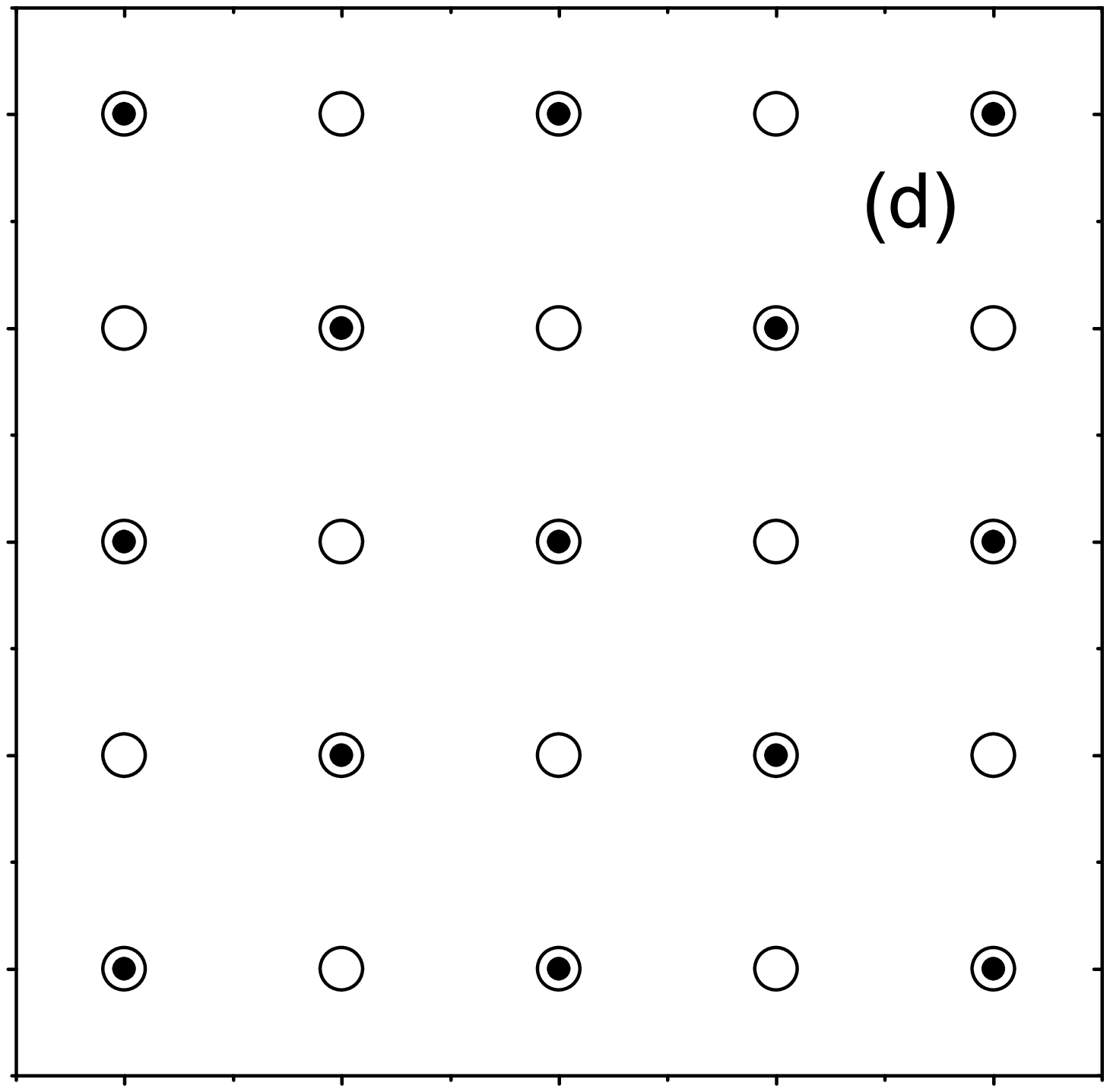}}
\text{Fig. 1(d)}
\text{Fig. 1(c)}
\epsfxsize= .85\hsize
\centerline{ \epsffile{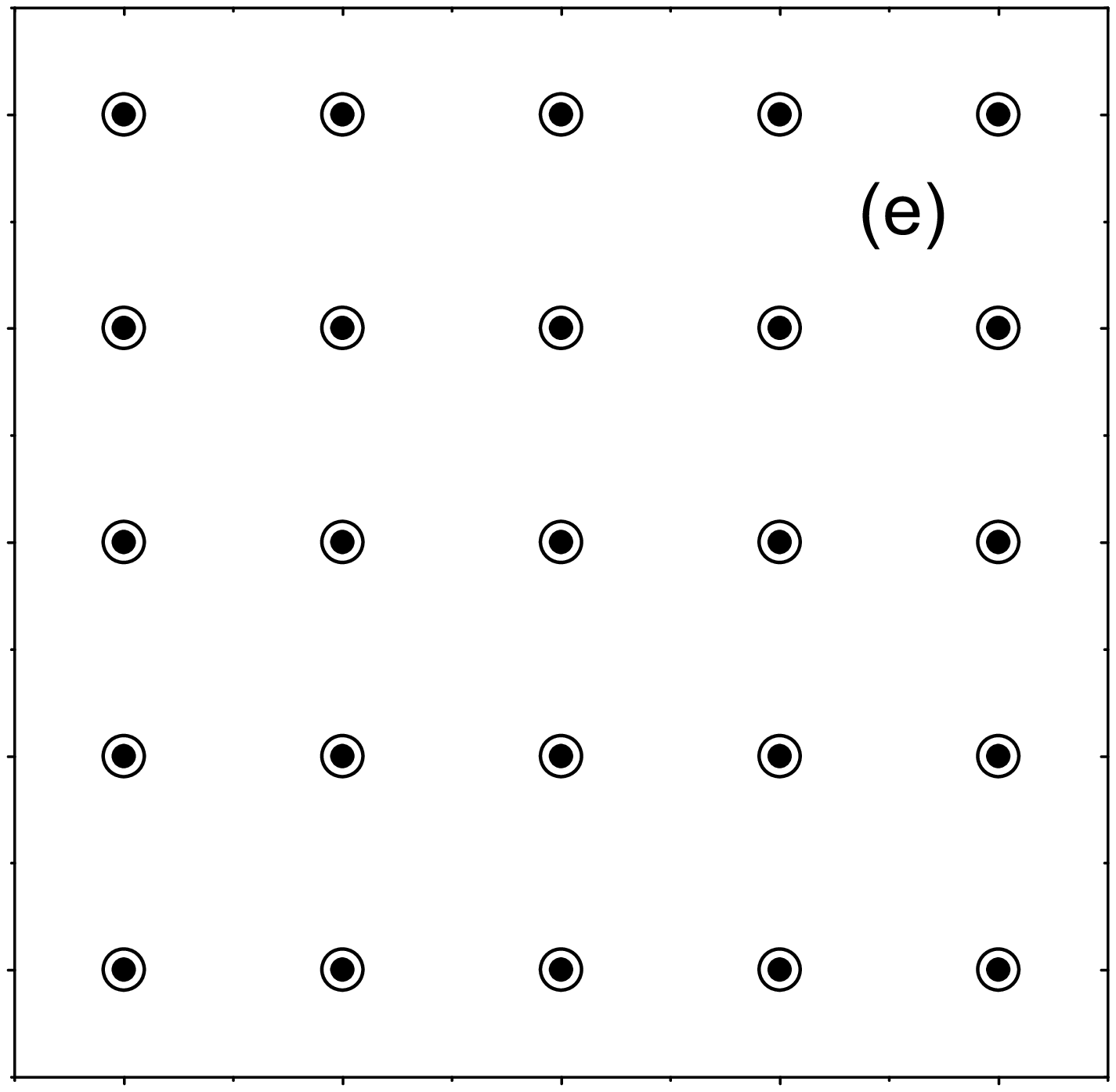}}
\text{Fig. 1(e)}
\end{figure}

\newpage
\begin{figure}[tbp]
\epsfxsize= .8\hsize
\centerline{ \epsffile{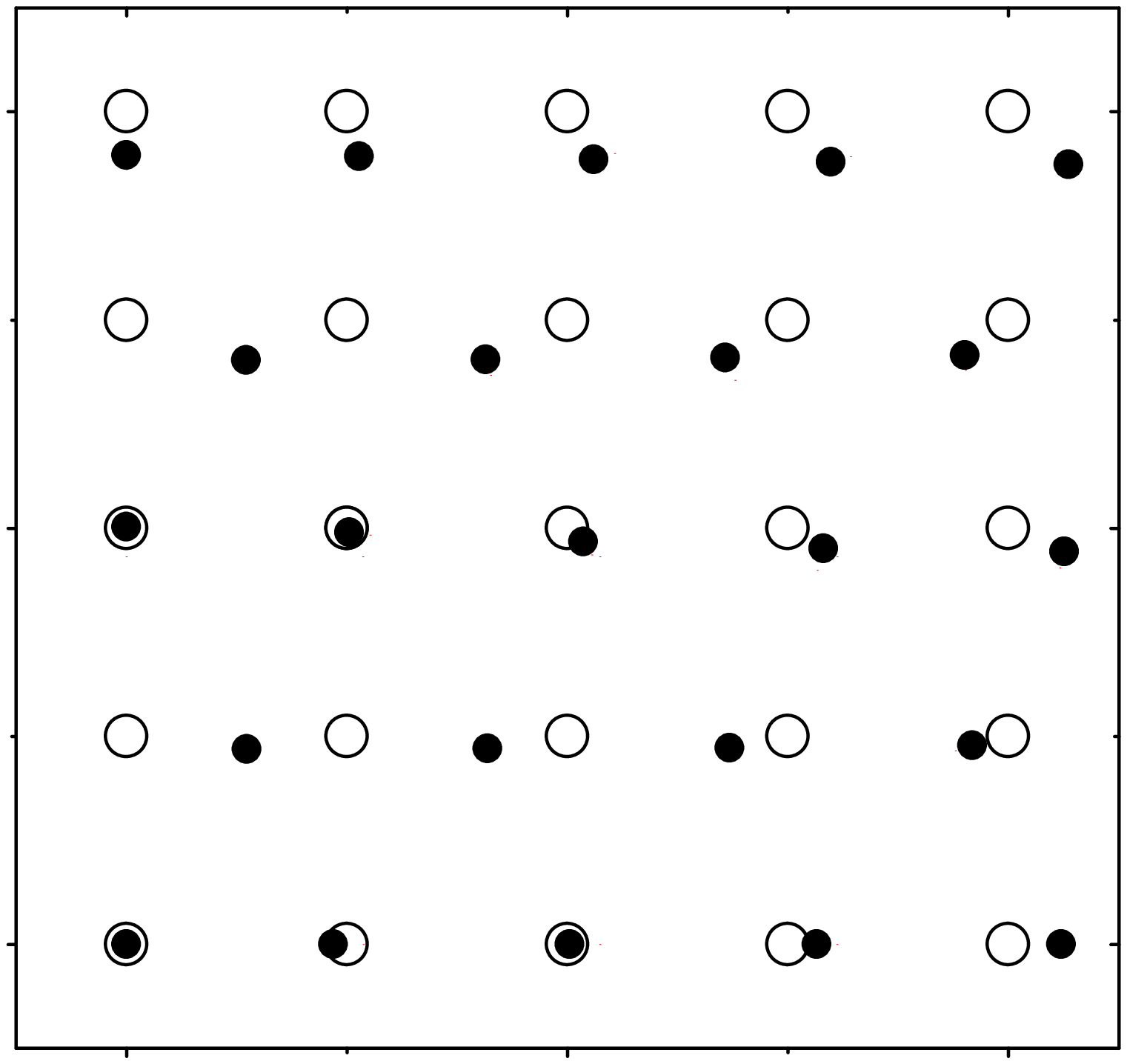}}
\text{Fig. 2}
\end{figure}
\begin{figure}[tbp]
\epsfxsize= .85\hsize
\centerline{ \epsffile{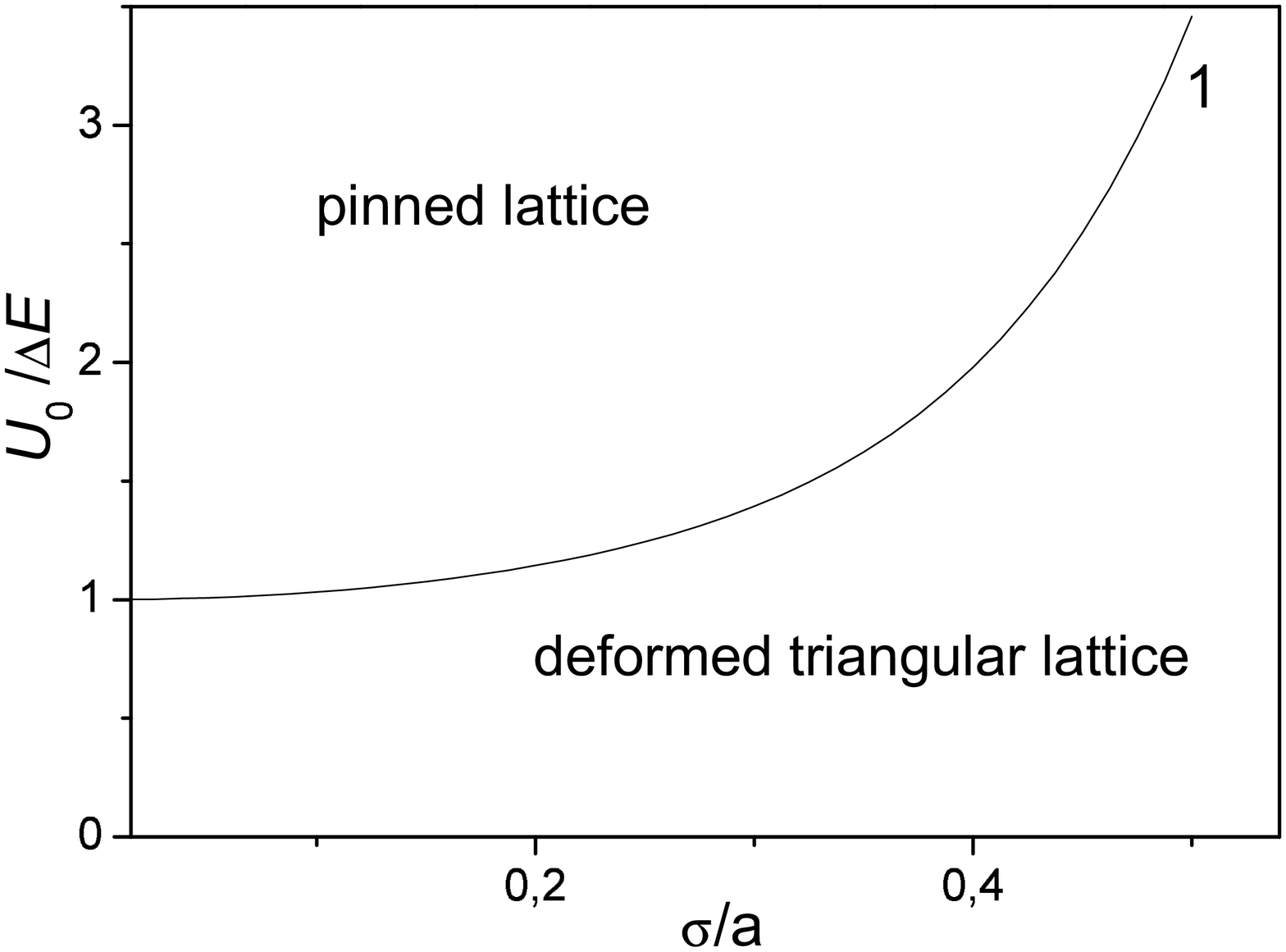}}
\text{Fig. 3}
\end{figure}
\newpage

\begin{figure}[tbp]
\epsfxsize= 0.85\hsize
\centerline{ \epsffile{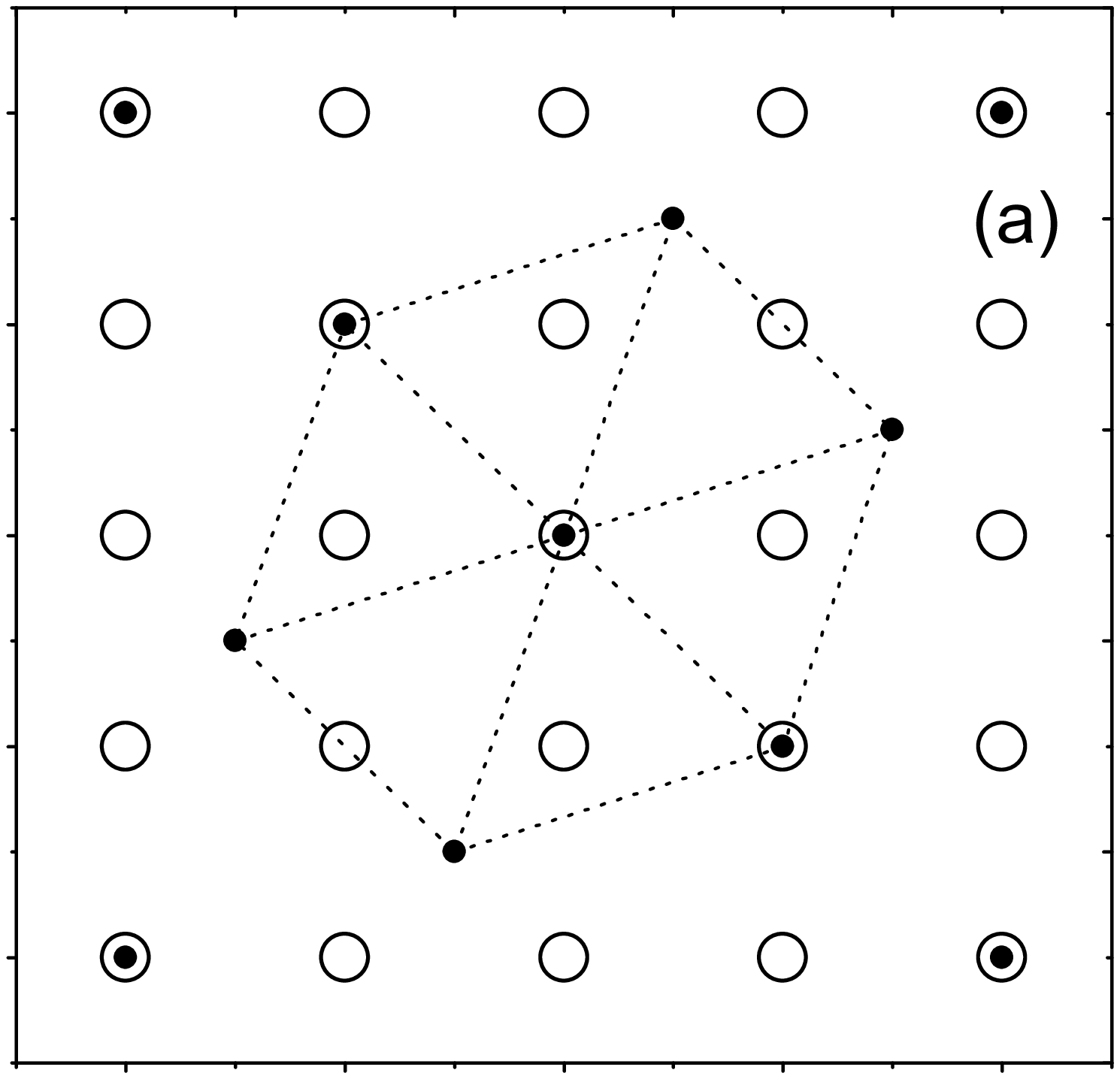}}
\text{Fig. 4(a)}
\epsfxsize= 0.85\hsize
\centerline{ \epsffile{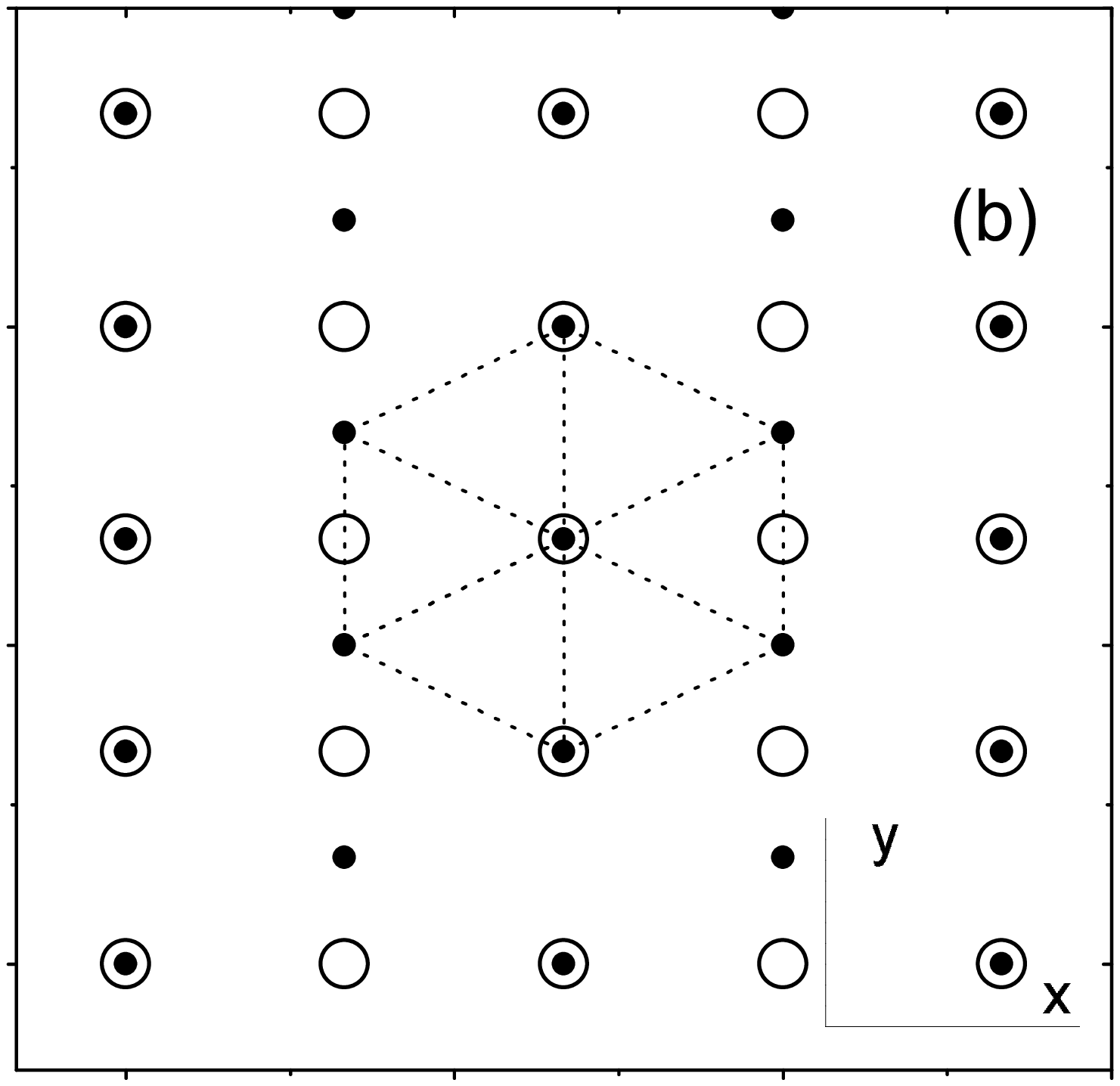}}
\text{Fig. 4(b)}
\end{figure}
\newpage

\begin{figure}[tbp]
\epsfxsize= 0.85\hsize
\centerline{ \epsffile{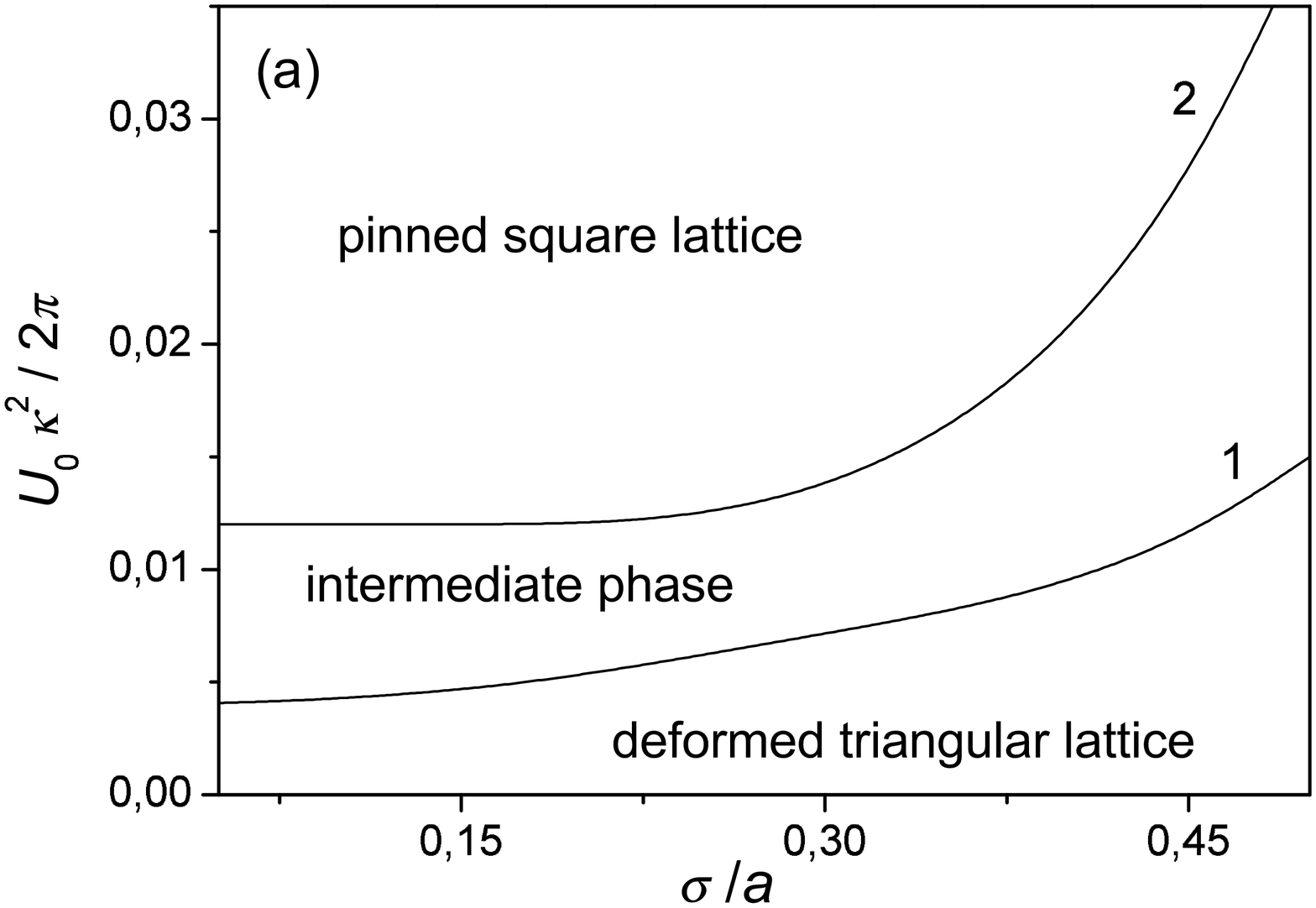}}
\text{Fig. 5(a)}
\epsfxsize= 0.85\hsize
\centerline{ \epsffile{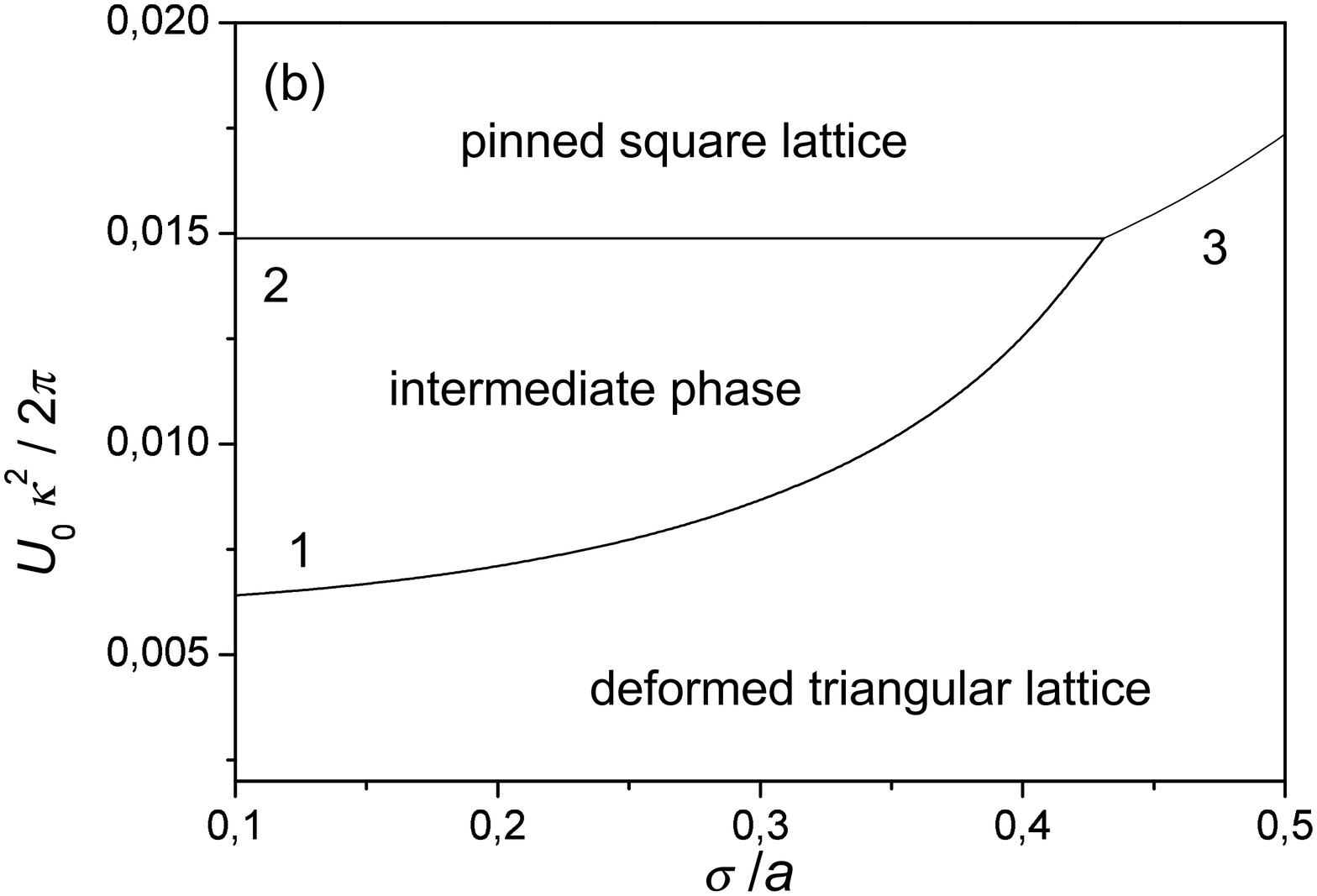}}
\text{Fig. 5(b)}
\end{figure}
\newpage

\begin{figure}[tbp]
\epsfxsize= .85\hsize
\centerline{ \epsffile{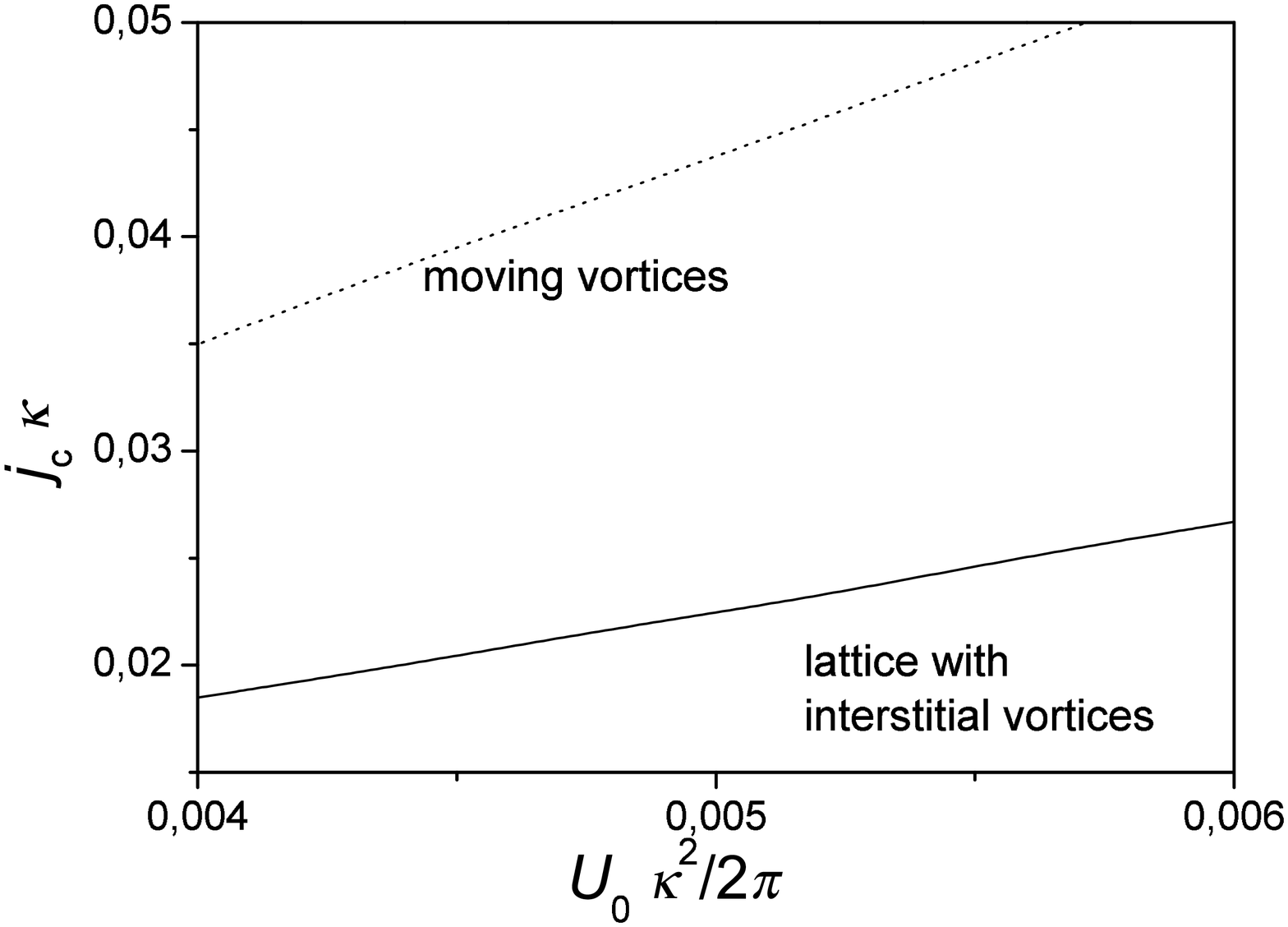}}
\text{Fig. 6}
\end{figure}

\begin{figure}[tbp]
\epsfxsize= .85\hsize
\centerline{ \epsffile{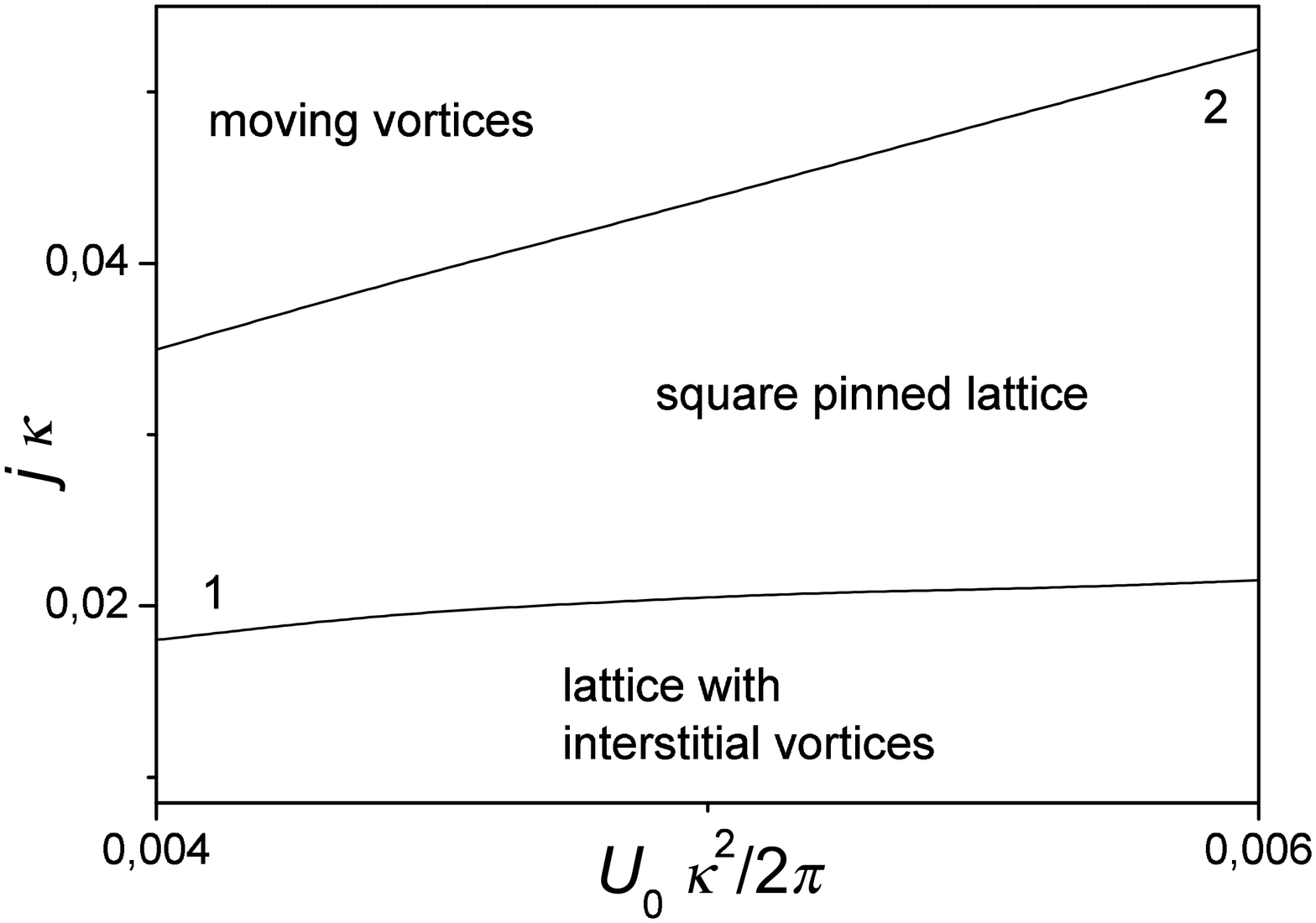}}
\text{Fig. 7}
\end{figure}

\end{document}